\documentclass[a4paper,11pt]{article}

\usepackage{contribution}

\newcommand{\weblink}[2][]{%
    \ifthenelse{\equal{#1}{}}%
    {\textnormal{\url{#2}}}%
    {\textnormal{\href{#2}{#1}}}%
}

\def\beq{\begin{equation}}
\def\eeq#1{\label{#1}\end{equation}}
\def\eeqn{\end{equation}}

\def\beqa{\begin{eqnarray}}
\def\eeqa#1{\label{#1}\end{eqnarray}}
\def\eeqan{\end{eqnarray}}

\let\bar=\overbar

\def\Dslash{\not{\hbox{\kern-4pt $D$}}}
\def\dslash{\not{\hbox{\kern-2pt $\del$}}}

\def\msb{{\bar{\ssstyle M \kern -1pt S}}}

\newcommand{\contribution}[7][]{%
  \clearpage
  \thispagestyle{plain}
  \ifthenelse{\equal{#1}{}}
  {\hypersetup{pdftitle={#2}}}
  {\hypersetup{pdftitle={#1}}}
  \hypersetup{pdfauthor={{#3} {#4}}}
  {\centering\normalfont\LARGE\bfseries\sffamily #2 \par\nobreak}
  \lhead{}
  \chead{%
    \textit{\footnotesize XIV International Conference on Hadron Spectroscopy
      (\weblink[\textit{hadron2011}]{http://www.hadron2011.de}), 13-17 June 2011, Munich, Germany}%
  }
  \rhead{}
  \bigskip
  \begin{center}
    {#3} {#4}\ifthenelse{\equal{#6}{}}{}{\footnote{\weblink[#6]{mailto:#6}}}
    \ifthenelse{\equal{#7}{}}{}{#7} \\
    \textit{#5}
  \end{center}
  \bigskip
}

\renewcommand{\abstract}[1]{%
  \begin{center}
    \begin{minipage}{0.85\textwidth}
      \begin{footnotesize}
        #1
      \end{footnotesize}
    \end{minipage}
  \end{center}
  \bigskip
}

\begin{document}

%
%
%
%
%
{  

\makeatletter
\@ifundefined{c@affiliation}%
{\newcounter{affiliation}}{}%
\makeatother
\newcommand{\affiliation}[2][]{\setcounter{affiliation}{#2}%
  \ensuremath{{^{\alph{affiliation}}}\text{#1}}}

\def\LamFOF{\Lambda (1405)}
\def\KbarN{\bar{K} N}

\def\Rho{\text{P}}
\def\PE{\Rho _{\text{E}}}
\def\PB{\Rho _{\text{B}}}
\def\PS{\Rho _{\text{S}}}

\def\fm{~\text{fm}}

%


%
\contribution[Internal structure of the $\Lambda (1405)$ resonance]
{Internal structure of the $\Lambda (1405)$ resonance
  probed in chiral unitary amplitude}
{T.}{Sekihara}  
{\affiliation[Department of Physics, Tokyo Institute of Technology, 
  Tokyo 152-8551, Japan]{1} \\
  \affiliation[Yukawa Institute for Theoretical Physics, 
  Kyoto University, Kyoto 606-8502, Japan]{2}}
{}
{\!\!\affiliation{1}, T. Hyodo\affiliation{1}, and D. Jido\affiliation{2}}


%

\abstract{%
  The internal structure of the resonant $\LamFOF$ state is
  investigated based on meson-baryon coupled-channels chiral dynamics.
  We evaluate $\LamFOF$ form factors which are extracted from
  current-coupled scattering amplitudes in meson-baryon degrees of
  freedom.
  Using several probe currents and channel decomposition, we find that
  the resonant $\LamFOF$ state is dominantly composed of widely spread
  $\bar{K}$ around $N$, with escaping $\pi \Sigma$ component.  }
%

\section{Introduction}

There are hadrons which are expected to have exotic structures, {\it
  e.g.}, hadronic molecules, and it is one of the important issues in
hadron physics to clarify their structures.  A classical example is
the excited baryon $\LamFOF$, which has been considered as an $s$-wave
$\KbarN$ quasibound state~\cite{Dalitz:1960du}.  It is also suggested
by the modern theoretical approach based on the chiral dynamics within
the unitary framework (the chiral unitary
approach)~\cite{Kaiser:1995eg,Oset:1997it,Oller:2000fj,Lutz:2001yb,Jido:2003cb,Hyodo:2011ur}
that $\LamFOF$ is dynamically generated in the meson-baryon
scattering, in addition to the good reproduction of the low-energy
$K^{-}p$ cross sections and the $\Lambda (1405)$ peak in $\pi \Sigma$
mass spectrum. Moreover, the chiral unitary approach predicts
double-pole structure for $\Lambda
(1405)$~\cite{Oller:2000fj,Jido:2003cb} and one of the poles is
expected to originate from the $\KbarN$ bound
state~\cite{Hyodo:2007jq,Hyodo:2008xr}. Some approaches for the survey
on the $\Lambda (1405)$ structure in experiments have been proposed,
e.g., in Refs.~\cite{Jido:2009jf,Cho:2010db}.

If $\Lambda (1405)$ is dominated by the $\KbarN$ quasibound state with
a small binding energy, one can expect that $\Lambda (1405)$ has a
larger size than typical ground state baryons dominated by genuine
$qqq$ components. Motivated by this expectation, in
Ref.~\cite{Sekihara:2008qk} we have investigated the internal
structure of the resonant $\Lambda (1405)$ state by evaluating density
distributions obtained from the form factors on the $\Lambda (1405)$
pole originating from the $\KbarN$ bound state.  In our study the
$\LamFOF$ form factors are directly extracted from the current-coupled
scattering amplitude, which involves a response of $\Lambda (1405)$ to
the external current. The current-coupled scattering amplitude is
evaluated in a charge conserved way by considering current couplings
to the constituent hadrons inside $\Lambda (1405)$.  Here we note that
the wave functions and form factors of $\Lambda (1405)$ were studied
also in Ref.~\cite{YamagataSekihara:2010pj} in a cut-off scheme within
chiral unitary approach, which results were not significantly
different from ours, except for the high momentum region compared to
the cut-off.

\section{Internal structure of $\Lambda (1405)$}

In the chiral unitary approach, the meson-baryon scattering amplitude
$T_{ij}$ with channel indices $i$ and $j$ is obtained by a
coupled-channels equation,
\begin{equation}
T_{ij} (\sqrt{s}) 
= V_{ij} (\sqrt{s}) 
+ \sum _{k} V_{ik} (\sqrt{s}) G_{k} (\sqrt{s}) T_{kj} (\sqrt{s}) , 
\end{equation}
with the interaction kernel $V_{ij}$ given by chiral perturbation
theory, a meson-baryon loop integral $G_{k}$, and the center-of-mass
energy $\sqrt{s}$.  The obtained amplitude contains dynamically
generated $\LamFOF$ in $s$ wave. Next, in order to observe response of
$\Lambda (1405)$ to the conserved probe current in the chiral unitary
approach, we evaluate current-coupled scattering amplitude $T_{\gamma
  ij}^{\mu}$ in a charge conserved way, considering current couplings
to the constituent hadrons as~\cite{Sekihara:2008qk,Borasoy:2005zg}:
\begin{equation}
T_{\gamma ij}^{\mu} (\sqrt{s^{\prime}}, \, \sqrt{s}; \, Q^{2} )
= T_{\gamma (1) ij}^{\mu} + T_{\gamma (2) ij}^{\mu} + T_{\gamma (3) ij}^{\mu} , 
\end{equation} 
with the squared current momentum $Q^{2}$ and 
\begin{equation}
T_{\gamma (1) ij}^{\mu} 
= \sum _{k} T_{ik} D_{\text{M}_{k}}^{\mu} T_{kj} , 
\quad 
T_{\gamma (2) ij}^{\mu} 
= \sum _{k} T_{ik} D_{\text{B}_{k}}^{\mu} T_{kj} , 
\quad 
T_{\gamma (3) ij}^{\mu} 
= \sum _{k, l} T_{ik} G_{k} \Gamma _{kl}^{\mu} G_{l} T_{lj} ,
\end{equation}
where $D_{\text{M}_{k}}$ and $D_{\text{B}_{k}}$ are respectively loop
integrals with the current couplings to the meson and baryon and
$\Gamma _{ij}$ represents $MBM^{\prime}B^{\prime}\gamma$ vertex.  Then
the $\Lambda (1405)$ form factor, $F^{\mu}(Q^{2})$, can be extracted
by~\cite{Sekihara:2008qk,Jido:2002yz},
\begin{equation}
F^{\mu} (Q^{2}) 
=  
- \frac{(\sqrt{s^{\prime}}-Z_{\text{R}})T_{\gamma ij}^{\mu}
(\sqrt{s^{\prime}}, \, \sqrt{s}; \, Q^{2} )}{T_{ij} (\sqrt{s})} 
\Bigg |_{\sqrt{s} \to Z_{\text{R}}}
\Bigg |_{\sqrt{s^{\prime}} \to Z_{\text{R}}} , 
\label{eq:Res_scheme}
\end{equation}
where $Z_{\text{R}}$ is the $\Lambda (1405)$ pole position.  Here we
note that we have following relations:
\begin{equation}
\hat{Q} \frac{d G_{k}}{d \sqrt{s}} = 
( D_{\text{M}_{k}}^0 + D_{\text{B}_{k}}^0 )|_{Q^{2}=0} , 
\quad 
\hat{Q} \frac{d V_{ij}}{d \sqrt{s}} = \Gamma _{ij}^0 |_{Q^{2}=0} , 
\label{eq:WardTakahashi}
\end{equation}
with $\hat{Q}$ being the charge of $\Lambda (1405)$ with respect to
the probe current.  These are the Ward-Takahashi identity for the
two-body free propagator $G_{k}$ and the elementary vertex $V_{ij}$.

\begin{figure}[!Ht]
  \begin{center}
    \begin{tabular*}{\textwidth}{@{\extracolsep{\fill}}cc}
      \includegraphics[width=0.48\textwidth]{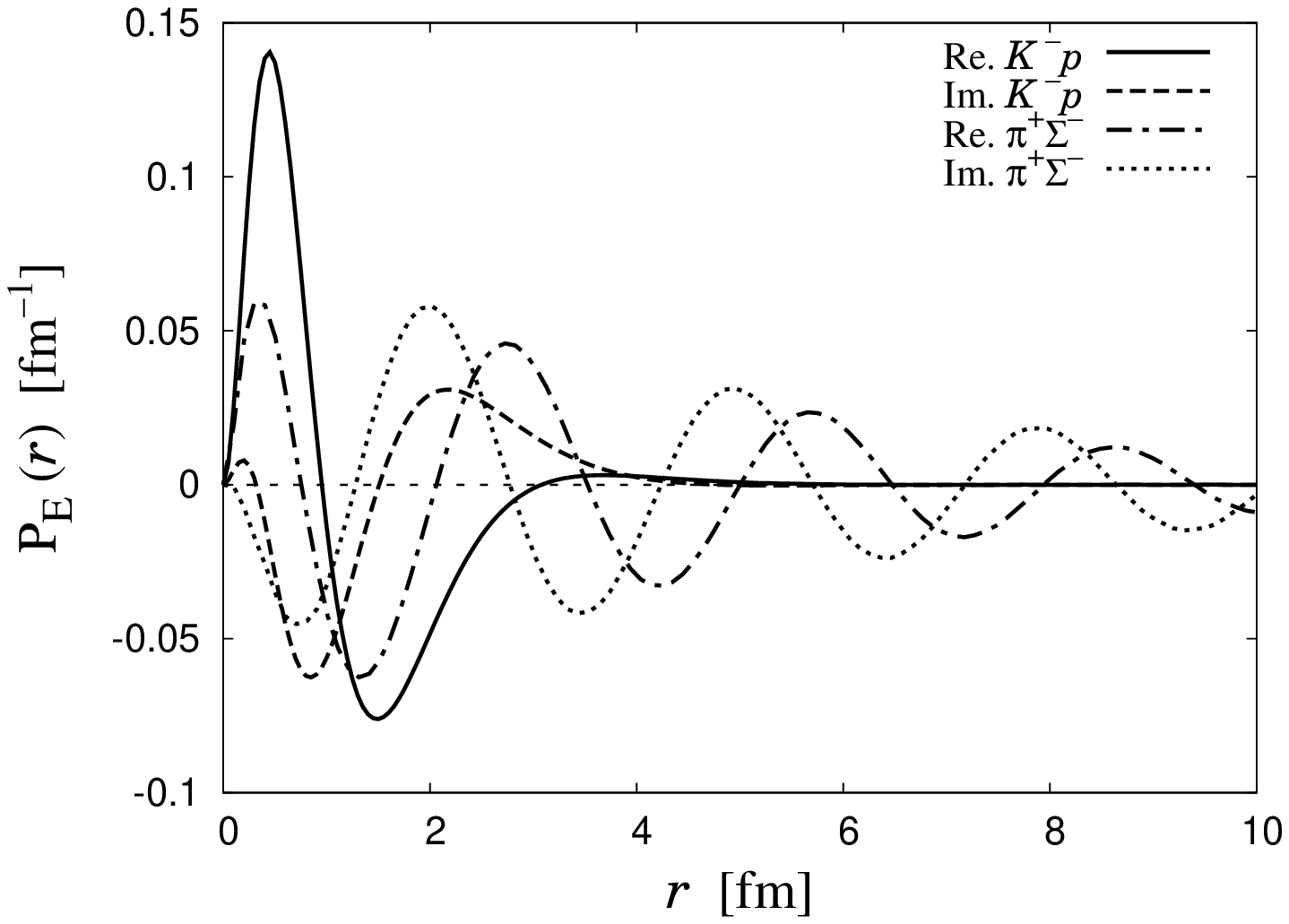} &
      \includegraphics[width=0.48\textwidth]{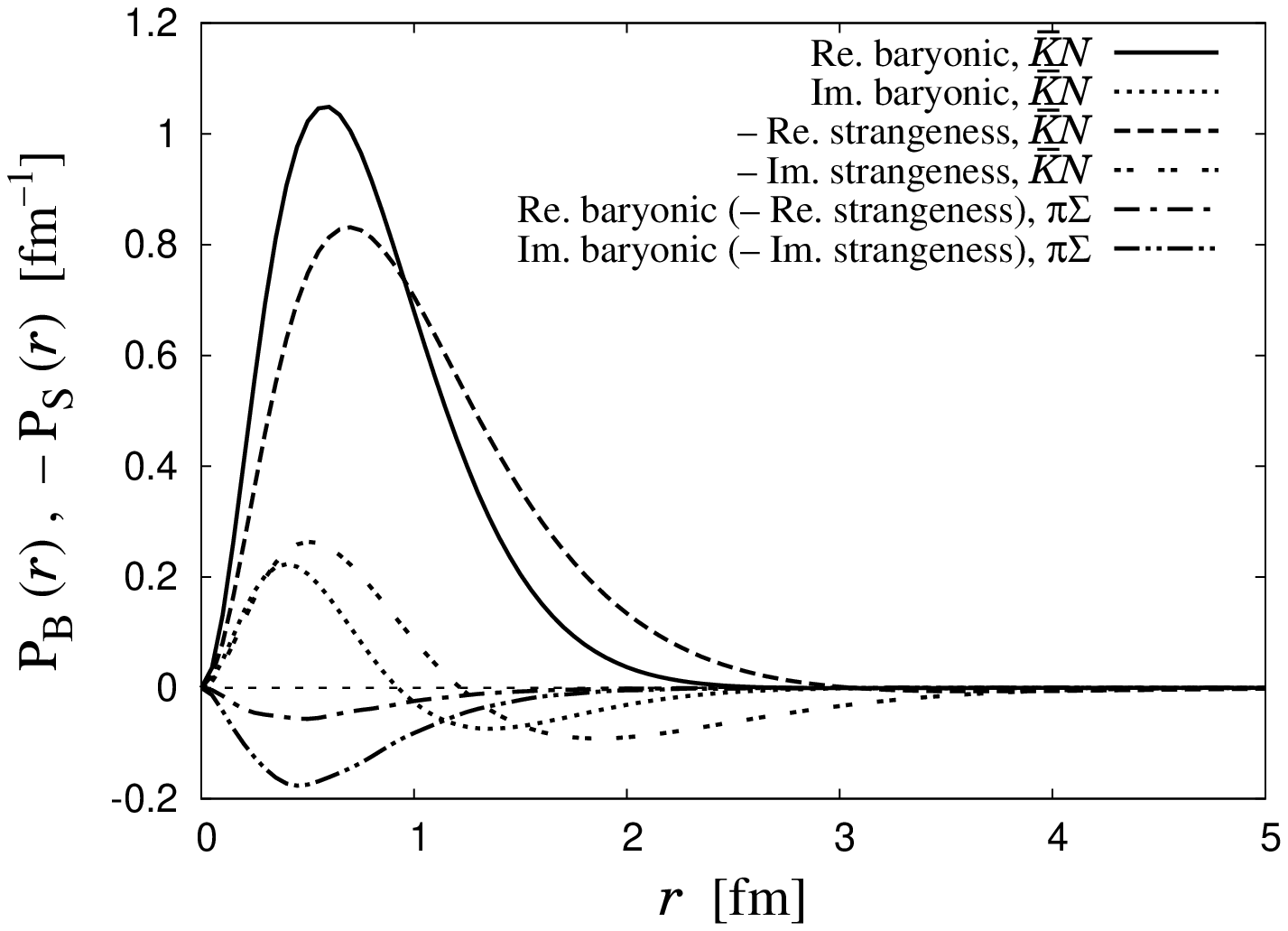}
    \end{tabular*}
    \caption{Electric ($\Rho _{\text{E}}$, left), baryonic and
      strangeness ($\Rho _{\text{B}}$ and $\Rho _{S}$, right) density
      distributions of $\Lambda (1405)$ in each component. Here $\Rho
      _{\text{E}}$ is shown in particle basis whereas $\Rho
      _{\text{B}}$ and $\Rho _{\text{S}}$ are in isospin
      basis~\cite{Sekihara:2008qk}. }
    \label{fig:Rho_pole}
  \end{center}
\end{figure}

Now let us show our results for the internal structure of the resonant
$\Lambda (1405)$.  First, we write a normalization relation for the
baryonic [$F_{\text{B}}(Q^{2})$] and strangeness
[$F_{\text{S}}(Q^{2})$] form factors of $\Lambda (1405)$ proved in
Ref.~\cite{Sekihara:2008qk},
\begin{equation}
F_{\text{B}}(Q^{2}=0) = - F_{\text{S}}(Q^{2}=0) = 
- \sum _{i, j} g_{i} g_{j} \left . 
\left ( \frac{d G_{i}}{d \sqrt{s}} \delta _{ij} 
+ G_{i} \frac{d V_{ij}}{d \sqrt{s}} G_{j} \right ) 
\right |_{\sqrt{s} \to Z_{\text{R}}} 
= 1 ,  
\label{eq:gg_dGds}
\end{equation}
where $g_{i} g_{j}$ is a residue of $T_{ij}$ at the $\Lambda (1405)$
pole position and $dG_{i}/d\sqrt{s}$ ($dV_{ij}/d\sqrt{s}$) term comes
from $D_{\text{M}_{i}}^{0}+D_{\text{B}_{i}}^{0}$ ($\Gamma _{ij}^{0}$)
at $Q^{2}=0$.  This relation corresponds to the Ward identity for the
vertex and wave-function renormalization factors, and this originates
from that we evaluate $T_{\gamma ij}^{\mu}$ in a charge conserved way
with current couplings satisfying Ward-Takahashi
identity~\eqref{eq:WardTakahashi}.  With this relation, we can pin
down the dominant component of the $\Lambda (1405)$ structure by
decomposing the summation in Eq.~\eqref{eq:gg_dGds}.  As a
result, we find that contribution from the $\KbarN (I=0)$ channel
($=-g_{\KbarN}^{2}dG_{\KbarN}/d\sqrt{s}$) is $0.994 + 0.048i$ whereas
contributions from other channels and the vertex term ($=-\sum
_{i,j}g_{i}G_{i}dV_{ij}/d\sqrt{s}G_{j}g_{j}$) are negligibly
small~\cite{Sekihara:2008qk}.  Therefore, this result indicates that
the $\bar{K} N (I=0)$ channel generates more than $99 \%$ of the
$\LamFOF$ charge, which is consistent with the $\KbarN$ quasibound
state picture for $\Lambda (1405)$.

Next we show the electric, baryonic, and opposite-sign strangeness
density distributions ($\PE$, $\PB$, and $-\PS$, respectively) of
$\LamFOF$ in each component in Fig.~\ref{fig:Rho_pole}.
From $\PE$, we can see that the negative (positive) charge
distribution appears in $\LamFOF$ due to the existence of lighter
$K^{-}$ (heavier $p$) in the outside (inside) region, bearing in mind
the $\KbarN$ dominance for $\LamFOF$.  Also it is interesting to see
the dumping oscillation in $\pi ^{+} \Sigma ^{-}$ (equivalently
$\pi^{-} \Sigma^{+}$ with the opposite sign) component in $\PE$ as the
decay of the system, although this is not observed in the total $\PE$
due to the cancellation of $\pi ^{+} \Sigma ^{-}$ and $\pi ^{-} \Sigma
^{+}$ components.
On the other hand, $\PB$ and $\PS$ indicate that inside $\LamFOF$ the
$\bar{K}$ component has longer tail than the $N$ component and
$\bar{K}$ distribution largely exceeds typical hadronic size $\lesssim
1 \fm$, bearing in mind that the baryonic (strangeness) current probes
the $N$ ($\bar{K}$) distribution inside $\LamFOF$.

\section{Summary}

We have investigated the internal structure of the resonant $\LamFOF$
state in the chiral unitary approach, in which $\LamFOF$ is
dynamically generated in meson-baryon coupled-channels chiral
dynamics.  Probing $\LamFOF$ with conserved current in a charge
conserved way, we have observed that $\KbarN$ component gives more
than $99 \%$ of the total $\LamFOF$ charge.  The electric density
distribution indicates that inside $\LamFOF$ lighter $K^{-}$ (heavier
$p$) exists in the outside (inside) region and the escaping $\pi
\Sigma$ component appears as the decay mode of $\LamFOF$.  Also from
the baryonic and strangeness density distributions we have found that
inside $\LamFOF$ the $\bar{K}$ component has longer tail than the $N$
component and $\bar{K}$ distribution largely exceeds typical hadronic
size $\lesssim 1 \fm$.

This work is partly supported by the Grand-in-Aid for Scientific
Research from MEXT and JSPS (No. 21840026, 22105507, 22740161, and
22-3389).



%

}  



\begin{thebibliography}{99}
  
\bibitem{Dalitz:1960du}
  R.~H.~Dalitz and S.~F.~Tuan,
  Annals Phys.\  {\bf 10}, 307 (1960).

\bibitem{Kaiser:1995eg}
  N.~Kaiser, P.~B.~Siegel and W.~Weise,
  Nucl.\ Phys.\  A {\bf 594}, 325 (1995).

\bibitem{Oset:1997it}
  E.~Oset and A.~Ramos,
  Nucl.\ Phys.\  A {\bf 635}, 99 (1998).

\bibitem{Oller:2000fj}
  J.~A.~Oller and U.~G.~Meissner,
  Phys.\ Lett.\  B {\bf 500}, 263 (2001).

\bibitem{Lutz:2001yb}
  M.~F.~M.~Lutz and E.~E.~Kolomeitsev,
  Nucl.\ Phys.\  A {\bf 700}, 193 (2002). 

\bibitem{Jido:2003cb}
  D.~Jido, J.~A.~Oller, E.~Oset, A.~Ramos and U.~G.~Meissner,
  Nucl.\ Phys.\  A {\bf 725}, 181 (2003). 

\bibitem{Hyodo:2011ur}
  T.~Hyodo and D.~Jido,
  Prog.\ Part.\ Nucl.\ Phys.\ (2011), 
  doi:10.1016/j.ppnp.2011.07.002. 

\bibitem{Hyodo:2007jq}
  T.~Hyodo and W.~Weise,
  Phys.\ Rev.\  C {\bf 77}, 035204 (2008). 

\bibitem{Hyodo:2008xr}
  T.~Hyodo, D.~Jido and A.~Hosaka,
  Phys.\ Rev.\  C {\bf 78}, 025203 (2008). 

\bibitem{Jido:2009jf}
  D.~Jido, E.~Oset and T.~Sekihara,
  Eur.\ Phys.\ J.\  A {\bf 42}, 257 (2009); 
%
  {\it ibid.} {\bf 47}, 42 (2011). 

\bibitem{Cho:2010db}
  S.~Cho {\it et al.}  [ExHIC Collaboration],
  Phys.\ Rev.\ Lett.\  {\bf 106}, 212001 (2011); 
  arXiv:1107.1302 [nucl-th].


\bibitem{Sekihara:2008qk}
  T.~Sekihara, T.~Hyodo and D.~Jido,
  Phys.\ Lett.\  B {\bf 669}, 133 (2008); 
%
  Phys.\ Rev.\  C {\bf 83}, 055202 (2011). 


\bibitem{YamagataSekihara:2010pj}
  J.~Yamagata-Sekihara, J.~Nieves and E.~Oset,
  Phys.\ Rev.\  D {\bf 83}, 014003 (2011). 

\bibitem{Borasoy:2005zg}
  B.~Borasoy, P.~C.~Bruns, U.~G.~Meissner and R.~Nissler,
  Phys.\ Rev.\  C {\bf 72}, 065201 (2005). 

\bibitem{Jido:2002yz}
  D.~Jido, A.~Hosaka, J.~C.~Nacher, E.~Oset and A.~Ramos,
  Phys.\ Rev.\  C {\bf 66}, 025203 (2002). 


\end{thebibliography}
\end{document}